\documentstyle[12pt,titlepage]{article}
\input epsfig.sty

\font\grande=cmr9.5 scaled \magstep4
\font\medio=cmr9.5 scaled \magstep2
\outer\def\beginsection#1\par{\medbreak\bigskip
      \message{#1}\leftline{\bf#1}\nobreak\medskip
\vskip-\parskip
      \noindent}

\def\laq{\raise 0.4ex\hbox{$<$}\kern -0.8em\lower 0.62
ex\hbox{$\sim$}}
\def\gaq{\raise 0.4ex\hbox{$>$}\kern -0.7em\lower 0.62
ex\hbox{$\sim$}}

\begin{document}
\bibliographystyle {unsrt}

\titlepage

\begin{flushright}
CERN-PH-TH/2007-143
\end{flushright}

\vspace{15mm}
\begin{center}
{\grande Time-dependent gravitating solitons}\\
\vspace{5mm}
{\grande in five dimensional warped space-times}\\
\vspace{15mm}
 Massimo Giovannini 
 \footnote{Electronic address: massimo.giovannini@cern.ch} \\
\vspace{6mm}

\vspace{0.3cm}
{{\sl Centro ``Enrico Fermi", Compendio del Viminale, Via 
Panisperna 89/A, 00184 Rome, Italy}}\\
\vspace{0.3cm}
{{\sl Department of Physics, Theory Division, CERN, 1211 Geneva 23, Switzerland}}
\vspace*{2cm}

\end{center}

\vskip 2cm
\centerline{\medio  Abstract}
Time-dependent soliton solutions are explicitly derived in a five-dimensional theory 
endowed with one (warped) extra-dimension. Some of the obtained geometries,
everywhere well defined and technically regular, smoothly interpolate between two 
five-dimensional anti-de Sitter space-times for fixed value of the conformal time coordinate. 
Time dependent solutions containing 
both topological and non-topological sectors are also obtained.
Supplementary degrees of freedom can be also included and, in this case, the resulting 
multi-soliton solutions may describe time-dependent kink-antikink systems.
\noindent

\vspace{5mm}

\vfill
\newpage

\renewcommand{\theequation}{1.\arabic{equation}}
\section{Physical motivations}
\setcounter{equation}{0}

Time-dependent soliton solutions arise typically in $(1+1)$ space-time dimensions by 
appropriately boosting static solutions. This observation is due to the pioneering 
work of various authors \cite{1,2,3} (see also \cite{4,5} and references therein 
for an extended introduction) whose main focus was on the Sine-Gordon system.
The static solutions of the $(1+1)$ dimensional system only involved one 
scalar degree of freedom. Furthermore, the geometry was always static and Minkowskian both in the case 
of static and time-dependent solutions. The boosted solutions could be simply interpreted, from the physical point 
of view, as traveling solitons which could even illustrate, with appropriate 
boundary conditions  (see, for instance, \cite{3}), the elastic collision of two profiles.

In $(1+1)$ dimensions, static
multisoliton solutions are also known to exist. The obtainable solutions can be categorized into two 
basic groups. To the first set belong solutions containing two (or more) topological defects.
The second group of solutions contemplates two (or more) defects of which at least 
one is a non-topological soliton. In the figurative language of some of the authors 
discussing these profiles they have been dubbed as bag-like solutions \cite{6,7,8}.
According to the same language it is also possible to achieve trapping-bag configurations where 
a topological kink is trapped by a non-topological bag (see also \cite{9,10} where trapping 
bag solutions appear in the framework of the so-called MIT-bag and SLAC-bag models of extended 
hadrons). 
The interest of multisoliton solutions in $(1+1)$ dimensions persists even in recent years 
and various interesting integration methods have been tailored  to deal, specifically, with 
non-linear multi-field equations \cite{11,12,13}. 

The main theme of the present paper will  not be $(1+1)$ dimensional systems
but rather five-dimensional gravitating soliton solutions. It is nonetheless inspiring to bear in mind the aforementioned 
observations since the topic of the present paper will be time-dependent generalizations 
of static soliton solutions in five-dimensional warped space-times.

Gravitating solitons in higher dimensions can be 
used to model, for instance, five-dimensional domain walls where the extra (bulk) coordinate 
is infinite and where the space-time geometry is consistently included 
as a solution of the appropriate higher-dimensional gravitational field equations. 
Single gravitating kinks in five-dimensions \cite{14,15,16,17,18,19,20} have been 
discussed (see also \cite{21,22}). Following the analogy suggested at the beginning of the 
present script, it is possible to find bag-like solutions in five-dimensional geometries 
with well defined $\mathrm{AdS}_{5}$ limit for large values of the bulk radius. 
It has been shown that multi-soliton solutions arise naturally also in a five-dimensional 
context and fully analytic solutions have been reported, always in warped geometries, 
where the field profiles describe both kink-antikink systems and trapping bag solutions \cite{22a}.

These discussions have been conducted, initially, in the 
framework of five-dimensional Einstein-Hilbert gravity. 
Possible extensions of this approach include
\begin{itemize}
\item{} the modification of the underlying gravitational theory (for instance from Einstein-Hilbert gravity 
to Gauss-Bonnet or Brans-Dicke gravity);
\item{} the increase of the dimensionality of the space-time (for instance from kinks in five dimensions
to Abelian vortices in six dimensions);
\item{} the increase in the number of defects (for instance from single-defect solutions 
to multi-defect solutions).
\end{itemize}
One of the distinctive aspects of the aforementioned attempts is that 
the obtained field profiles are thick, i.e. they have an internal structure and they 
are obtained as a consistent solution of the underlying gravitational theory. In this 
sense the nature of the physical system greatly differs from their
$(1+1)$-dimensional counterpart. Particularly interesting situations  are the ones 
where the geometry is completely regular \footnote{In this paper, when talking about 
regular geometry, we will always refer to the case when the relevant curvature invariants 
are finite for all values of the coordinates lying in the domain of definition of the 
solutions.} and it has well defined $\mathrm{AdS}_{5}$ limits for large absolute value of the bulk radius. 

The gravity theory employed in the actual constructions plays a relevant role in the physical properties 
of the solutions. As already mentioned, the obvious choice is to embed the defects 
in a higher-dimensional Einstein-Hilbert gravity theory (see, for instance, 
\cite{20} and references therein). However, also different 
choices are equally plausible. For instance one might choose to select a theory where the Einstein-Hilbert 
terms is non-minimally coupled to the scalar degree of freedom \cite{23,24,25,26}. 
Gravitating kinks have been also studied in quadratic gravity theories where the 
Einstein-Hilbert action is complemented by the Gauss-Bonnet combination \cite{27,28,29,29a}.
It is finally possible to combine these two options and discuss the possibility of higher-dimensional
solutions where the Einstein-Hilbert action is complemented by the Gauss-Bonnet combination 
and the dilaton couples explicitly to the curvature in the string frame metric. This possibility 
has been developed in \cite{30,31,31a}.

Another arena for thick defects involves also
the increase of the dimensionality of the space-time. 
In the framework of the Abelian-Higgs model it is possible to find thick string solutions 
in a six-dimensional space-time with well defined $\mathrm{AdS}_{6}$ limit 
far from the vortex and a Minkowskian limit close to the core of the defect. These solutions 
may arise, with different features, both within Einstein-Hilbert gravity \cite{32,32a,33,33a,34} and within 
Gauss-Bonnet gravity \cite{35}. Another example is represented by higher dimensional
hedgehogs which are present in the case of seven (warped) dimensions (see \cite{36,37} and references 
therein). Finally, there exist, always in a five-dimensional context, solutions 
where the solitons have non-topological properties \cite{38}, i.e. the static fields have a bag-like profile 
as a function of the bulk radius.

In the present paper the main theme will be to investigate time-dependent soliton solutions 
that are inspired by the symmetries of the static solutions. There is the hope, in this context, of finding 
new classes of cosmological models that do not follow in the class of separable 
solutions that have been already exploited in the five-dimensional case. The latter solutions 
are typically obtained in the presence of thin brane configurations that arise as 
Dirac delta functions in the energy-momentum tensor of the brane. Typically these solutions 
are singular at the location of the brane and the defects, being thin, do not have internal structure. 
In different words we can say the the interest of the present exercises resides mainly in the 
fact that time-dependent solitonic solutions can be eventually used to describe, for a fixed value 
of the bulk radius, higher-dimensional cosmological models which are, really and truly, stemming 
from thick gravitating defects or, for short, thick branes.

To introduce the discussion in more specific terms consider the static solutions of the following 
five-dimensional theory \footnote{In Eq. (\ref{action}) $\kappa = 8\pi G_{5}= 8\pi/M^{3}$ and $G_{5}$ is the 
five-dimensional Newton constant.}:
\begin{equation}
S = \int d^{5} x \sqrt{|G|} \biggl[ - \frac{R}{2 \kappa} + \frac{1}{2} G^{\mathrm{AB}} \partial_{A}\varphi
\partial_{B}\varphi - V(\varphi) \biggr],
\label{action}
\end{equation}
in the case of the line element 
\begin{equation}
ds^2 = G_{AB} dx^{A} dx^{B} = a^2(w) [\eta_{\mu\nu} dx^{\mu} dx^{\nu} - dw^2],
\label{LEL1}
\end{equation}
where $w$ is the bulk coordinate, $a(w)$ is the (static) warp factor and $\eta_{\mu\nu}$ is the four-dimensional Minkowski metric 
\footnote{Conventions: capital Latin indices run over the whole dimensionality of the space-time 
while Greek indices run over the four-dimensional space-time; Latin (lowercase) indices run 
over the three-dimensional spatial geometry. The signature of the metric 
is mostly minus, i.e. $(+,-,-,-,-)$}.
Consider then, a specific solution of this theory (see \cite{14,16,18} and references therein), i.e. 
\begin{equation}
a(w) = \frac{1}{\sqrt{\lambda^2 w^2 +1}}, \qquad \varphi = \sqrt{\frac{3}{\kappa}} \arctan{(b w)},
\label{SOL1}
\end{equation}
arising when the potential $V(\phi)$ takes the following trigonometric form:
\begin{equation}
V(\varphi) = \frac{3 \lambda^2}{2\kappa} \biggl[ 1 - 5 \sin^2{\Phi}\biggr],\qquad \Phi = \sqrt{\frac{3}{\kappa}} \varphi,
\label{SOL2}
\end{equation}
From Eqs. (\ref{SOL1}) and (\ref{SOL2}) we can infer two kinds of solutions:
\begin{itemize}
\item{} boosted solutions that can be obtained by replacing $w$ with $\gamma(u) ( w + u t)$ or 
with $\gamma(u) ( w - u t)$ (where $\gamma(u)= (1 - u^2)^{-1/2}$);
\item{} time-shifted solutions.
\end{itemize}
The first class of solutions describes kinks that travel along 
the fifth coordinate and they will be discussed a bit later in this paper. The second class of solutions 
can be more directly interpreted as a time-dependent cosmological model for a fixed value of the 
bulk coordinate (that can be chosen to be, for instance, at $w=0$). The trouble with this second 
class of solutions is that, in general, they will not be diagonal. Indeed, from Eqs. (\ref{LEL1}), (\ref{SOL1}) 
and (\ref{SOL2}) it is possible to deduce that a time-dependent solution of the theory can be found, always 
for the potential (\ref{SOL2}), for a line element of the type \footnote{In what follows the time 
coordinate shall be denoted by $\tau$. This is motivated by the fact, as it will be shown, that 
the geometries written in the parametrization adopted in the present script reduce to what we would call 
a conformally flat metric for a {\em fixed} value of the bulk coordinate. In this case 
the time coordinate, i.e. $\tau$ will be what is customarily called, in cosmological applications, 
conformal time coordinate (as opposed to the cosmic time coordinate where all the clocks are 
synchronized). This is the reason why we prefer to use $\tau$ rather than $t$ to denote the time.}
\begin{equation}
ds^2 = a^2(w,\tau) [ d \tau^2 - d\vec{x}^2 - d w^2 - 2 d w\,d \tau],
\label{LEL2}
\end{equation}
where 
\begin{equation}
a(w,t)= \frac{1}{\sqrt{\lambda^2 (w+\tau)^2 +1}}, \qquad \varphi = \sqrt{\frac{3}{\kappa}} \arctan{[\lambda (w+ \tau)]},
\label{SOL1a}
\end{equation}
The easiest way of getting convinced of this statement is by using the symmetry of the metric and then, as 
a cross-check, to verify that the transformed metric is a solution of the action (\ref{action}) with the potential 
given in Eq. (\ref{SOL2}). Indeed, the Ricci tensor obtained from Eqs. (\ref{LEL2}) and (\ref{SOL1a}) 
has covariant components that can be computed to be:
\begin{eqnarray}
R_{00} &=& 2 \lambda^2 \frac{1 + 2 \lambda^2 (w +\tau)^2}{[ 1 + \lambda^2 (w +\tau)^2]^2},
\nonumber\\
R_{ij}  &=& \lambda^2 \frac{1 - 4 \lambda^2 (w + \tau)^2}{[ 1 + \lambda^2 (w +\tau)^2]^2}\,\delta_{ij},
\nonumber\\
R_{ww} &=& 4 \lambda^2 \frac{1 -  \lambda^2 (w + \tau)^2}{[ 1 + \lambda^2 (w +\tau)^2]^2},
\nonumber\\
R_{w0} &=& R_{0w} = 4 \lambda^2 \frac{1 -  \lambda^2 (w + \tau)^2}{[ 1 + \lambda^2 (w +\tau)^2]^2}.
\label{ricciexample1}
\end{eqnarray}
By contraction of the Einstein equations derived from Eq. (\ref{action}), the relevant system 
can be recast in the following form:
\begin{eqnarray}
&& R_{A B} = \kappa\biggl[ \partial_{A}\varphi \partial_{B} \varphi - \frac{2}{3} V\biggr],
\label{EQ1}\\
&& G^{A B} \nabla_{A} \nabla_{B} \varphi + \frac{\partial V}{\partial\varphi} =0,
\label{EQ2}
\end{eqnarray}
where $\nabla_{A}$ is the covariant derivative computed in terms of the five-dimensional 
metric.  Using the field profile appearing in Eq. (\ref{SOL1a}) and the potential of Eq. (\ref{SOL2}) into Eqs. 
(\ref{EQ1}) and (\ref{EQ2}) the system is satisfied and the solution is proven to exist. 
Notice that the off-diagonal form of the metric reported in Eq. (\ref{LEL2}) is essential to match the off-diagonal 
components of the Ricci tensor (i.e. $R_{0w}$ and $R_{w0}$) which are now non-vanishing 
since the warp factor depends explicitly upon time. Indeed, from Eq. (\ref{LEL2}), the $0w$ and $w0$ 
components of Eq. (\ref{EQ1}) lead also to a generalized momentum constraint  that mixes 
derivatives of the field profile with respect to $w$ and with respect to $\tau$. The momentum 
constraint is solved iff the metric is in the form of Eq. (\ref{LEL2}), i.e. off-diagonal.

This example is, at the same time, tangential and inspiring. It is tangential since, as 
already remarked, the obtained geometry is necessarily off-diagonal and, therefore, difficult 
to interpret as a cosmological metric at fixed value of the bulk coordinate. It is, however, also 
inspiring since it suggests that it might not be hopeless to find time-dependent 
soliton solutions leading to a non-separable form of the warp factor. By separable form 
of the warp factor we mean solutions where $a(w,\tau)$ can be written as the product 
of two functions one depending solely upon $w$ and the other depending solely upon $\tau$.
Separable solutions are normally discussed in the framework of thin brane models 
but, as we will argue, they are rather unnatural in the case of thick defects.

The plan of the forthcoming discussion will therefore be the following. In section 2 we will look 
for diagonal (time-dependent and non separable) solutions in the five-dimensional case and with 
a single scalar profile. In section 3 this discussion will be generalized to the case of multi-defects.
Section 4 contains some complementary material on traveling solitons.
Finally section 5 contains our concluding remarks. 

\renewcommand{\theequation}{2.\arabic{equation}}
\section{Time-dependent gravitating defects}
\setcounter{equation}{0}

To improve on the off-diagonal ansatz shown to satisfy the whole system 
of equations supported by time-dependent (gravitating) profiles discussed 
in the introduction, it is appropriate to consider the following ansatz:
\begin{equation}
ds^2 = G_{AB} d x^{A} dx^{B} = a^2(w,\tau)[d\tau^2 - d\vec{x}^2] - b^2(w,\tau) dw^2.
\label{LEL3}
\end{equation}
Using Eq. (\ref{LEL3}) for the calculation of the Ricci tensor and of the Christoffel 
connections the various components of Eq. (\ref{EQ1}) easily become, after some algebra:
\begin{eqnarray}
&& \frac{a^2}{b^2} [ {\mathcal H}' + 4 {\mathcal H}^2 - {\mathcal H} {\mathcal F}] 
- [ \dot{F} + 3 \dot{H} + F( F - H)]
= \kappa \biggl[ \dot{\varphi}^2 - \frac{2}{3} V a^2\biggr],
\label{EQ100}\\
&& - \frac{a^2}{b^2} [ {\mathcal H}' + 4 {\mathcal H}^2 - {\mathcal H} {\mathcal F}]
+ [\dot{H} + 2 H^2 + HF] = \frac{2}{3} \kappa V a^2,
\label{EQ1ij}\\
&& \frac{b^2}{a^2} [\dot{F} +F^2 + 2 H F] - 4 {\mathcal H}' + 4 {\mathcal H} ({\mathcal F} - {\mathcal H}) = 
\kappa \biggl[ {\varphi'}^2 + \frac{2}{3} V b^2\biggr],
\label{EQ1ww}\\
&& \dot{\mathcal H} - 4 H' + 3 F {\mathcal H} = \kappa \dot{\varphi} \varphi'.
\label{EQ10w}
\end{eqnarray}
The notations employed in Eqs. (\ref{EQ100}), (\ref{EQ1ij}), (\ref{EQ1ww}) and (\ref{EQ10w}) are the following:
\begin{itemize}
\item{} the prime denotes a derivation with respect to the bulk radius $w$;
\item{} the over-dot denotes a derivation with respect to the conformal time coordinate $\tau$;
\item{} ${\mathcal H}$ and ${\mathcal F}$ are related with the derivatives of $a$ and $b$ with respect to $w$, 
i.e. more specifically,
\begin{equation}
{\mathcal H}(w,\tau) = \frac{\partial \ln{a}}{\partial w},\qquad {\mathcal F}(w,\tau) = \frac{\partial \ln{b}}{\partial w};
\label{HFw}
\end{equation}
\item{}$H$ and $F$ are related with the derivatives of $a$ and $b$ with respect to $\tau$, 
i.e. more specifically,
\begin{equation}
H(w,\tau) = \frac{\partial \ln{a}}{\partial \tau},\qquad  F(w,\tau) = \frac{\partial \ln{b}}{\partial \tau}.
\label{HFtau}
\end{equation}
\end{itemize}
Equations (\ref{EQ100}), (\ref{EQ1ij}) and (\ref{EQ1ww}) follow, respectively, from 
the $(00)$, $(i=j)$ and $(ww)$ components of Eq. (\ref{EQ1}) while Eq. (\ref{EQ10w}) 
follows from the $(0w)$ component of Eq. (\ref{EQ1}). 
Finally, Eq. (\ref{EQ2}) leads, within the notations expressed by Eqs. (\ref{HFw}) and (\ref{HFtau}), 
to the following explicit relation
\begin{equation}
\ddot{\varphi} - \frac{a^2}{b^2}\varphi'' + (2 H + F) \dot{\varphi} - \frac{a^2}{b^2}( 4 {\mathcal H} - {\mathcal F}) \varphi' + \frac{\partial V}{\partial\varphi}a^2=0.
\label{KG1}
\end{equation}
If the dependence upon $\tau$ and $w$ would be separable (i.e. $a(w,\tau) = \alpha(\tau) \tilde{a}(w)$ 
and $b(w,\tau) = \beta(\tau) \tilde{b}(w)$), then, necessarily, ${\mathcal H}$ and ${\mathcal F}$ 
will only be function of the bulk radius (i.e.
${\mathcal H}(w)$ and ${\mathcal F}(w)$), while $H$ and $F$ will only be functions 
of the conformal time coordinate (i.e. 
$H(\tau)$ and $F(\tau)$). The constraint of Eq. (\ref{EQ10w}) can then be trivially 
satisfied by requiring either $F=0$ (i.e. $\beta $ constant in $\tau$)
  or ${\mathcal H} =0$ (i.e. $\tilde{a}$ constant in $w$). This situation is, however, not generic.
  
  It is appropriate to mention that by taking linear combinations of some of the 
  above equations, the resulting expressions can be put in a more friendly form. In particular, by summing up 
  Eqs. (\ref{EQ100}) and (\ref{EQ1ij}) we do get:
  \begin{equation}
  \dot{\varphi}^2 = \frac{1}{\kappa} [ 2 H^2 + 2 HF - F^2 - \dot{F} - 2 \dot{H}].
  \label{COMB1}
  \end{equation}
  By combining Eqs. (\ref{EQ1ij}) and (\ref{EQ1ww}) we do get instead
  \begin{equation}
  {\varphi'}^2 = \frac{3}{\kappa}( {\mathcal H} {\mathcal F} - {\mathcal H}') + \frac{1}{\kappa} \frac{a^2}{b^2}[(\dot{F} - \dot{H})  + (F + 2 H) ( F - H)].
  \label{COMB2}
  \end{equation}
 Consider then the following form of the warp factors:
 \begin{equation}
 a(w,\tau) = \frac{1}{\sqrt{\lambda^2 ( w + \tau)^2 + 1}}, \qquad  
 b(w,\tau) = \frac{\epsilon}{\sqrt{\lambda^2 ( w + \tau)^2 + 1}},
 \label{S1}
\end{equation}
where $\epsilon$ is both constant in $w$ and in $\tau$. It can be verified that 
Eq. (\ref{S1}) satisfies the whole system of equations provided the scalar field 
and the potential are:
\begin{eqnarray}
&& \varphi (w,\tau) = \sqrt{\frac{3}{\kappa}} \arctan{[\lambda( w + \tau)]}, \qquad 
\label{S2}\\
&& V(\varphi) = \frac{3 \lambda^2}{ 2 \kappa} \frac{\epsilon^2 -1}{\epsilon^2} ( 5 \sin^2{\Phi} -1), \qquad 
\Phi = \sqrt{\frac{\kappa}{3}} \varphi.
 \label{S3}
 \end{eqnarray}
If $\epsilon \to 1$ the solution does not reduce to a static limit but rather it reduces to a free solution 
with vanishing potential (i.e. $V(\varphi) =0$) where the explicit expression of $\varphi$ is still 
given by Eq. (\ref{S2}). Again, the solution $\epsilon =1$ does not have a static analog since the time dependence 
is essential. In other words $\Phi(w) = \arctan{\lambda w}$ is not a solution of the system if we 
set the time dependence to zero and $V(\Phi) =0$ from the very beginning. 
\begin{figure}
\begin{center}
\begin{tabular}{|c|c|}
      \hline
      \hbox{\epsfxsize = 7.5 cm  \epsffile{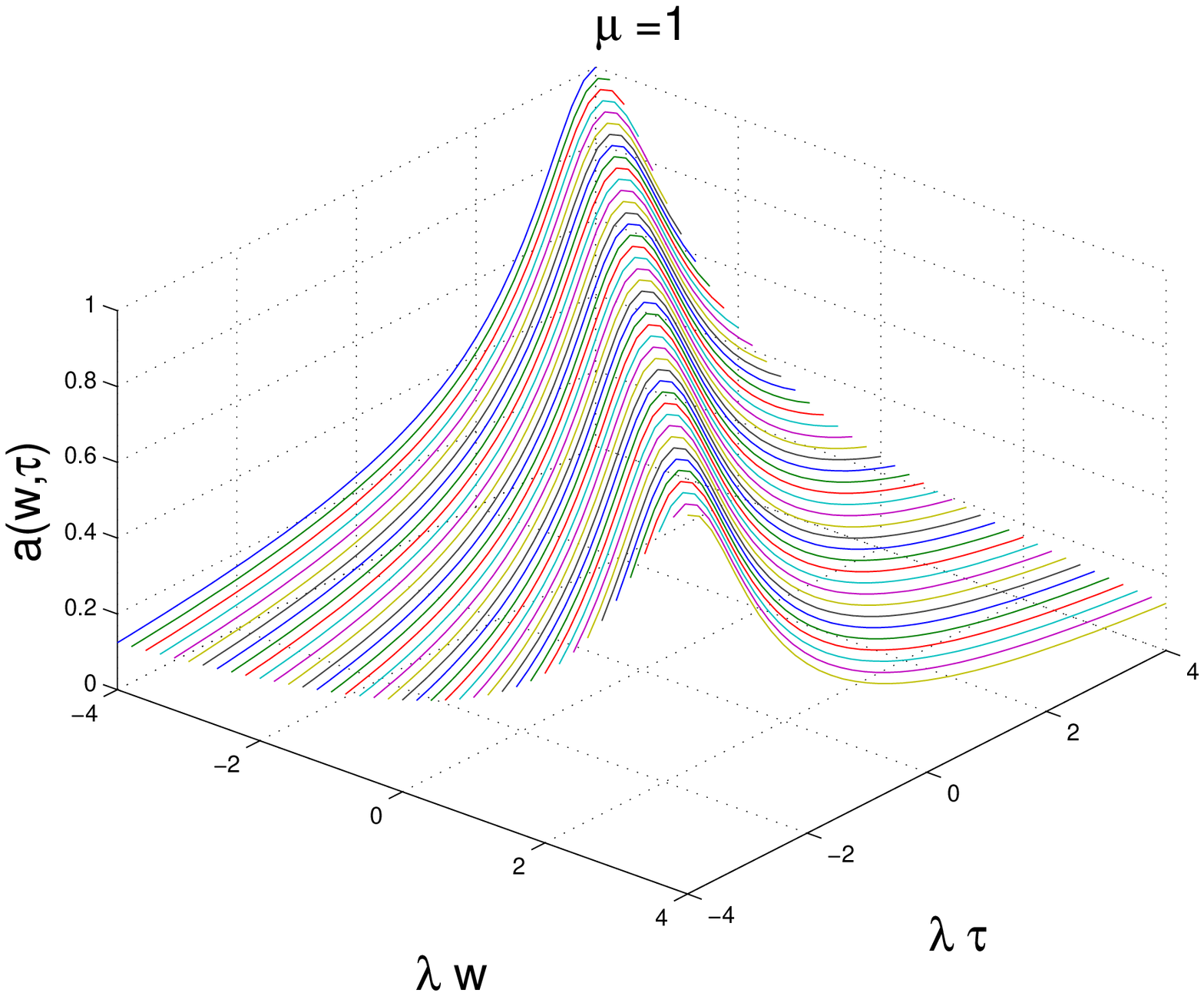}} &
      \hbox{\epsfxsize = 7.5 cm  \epsffile{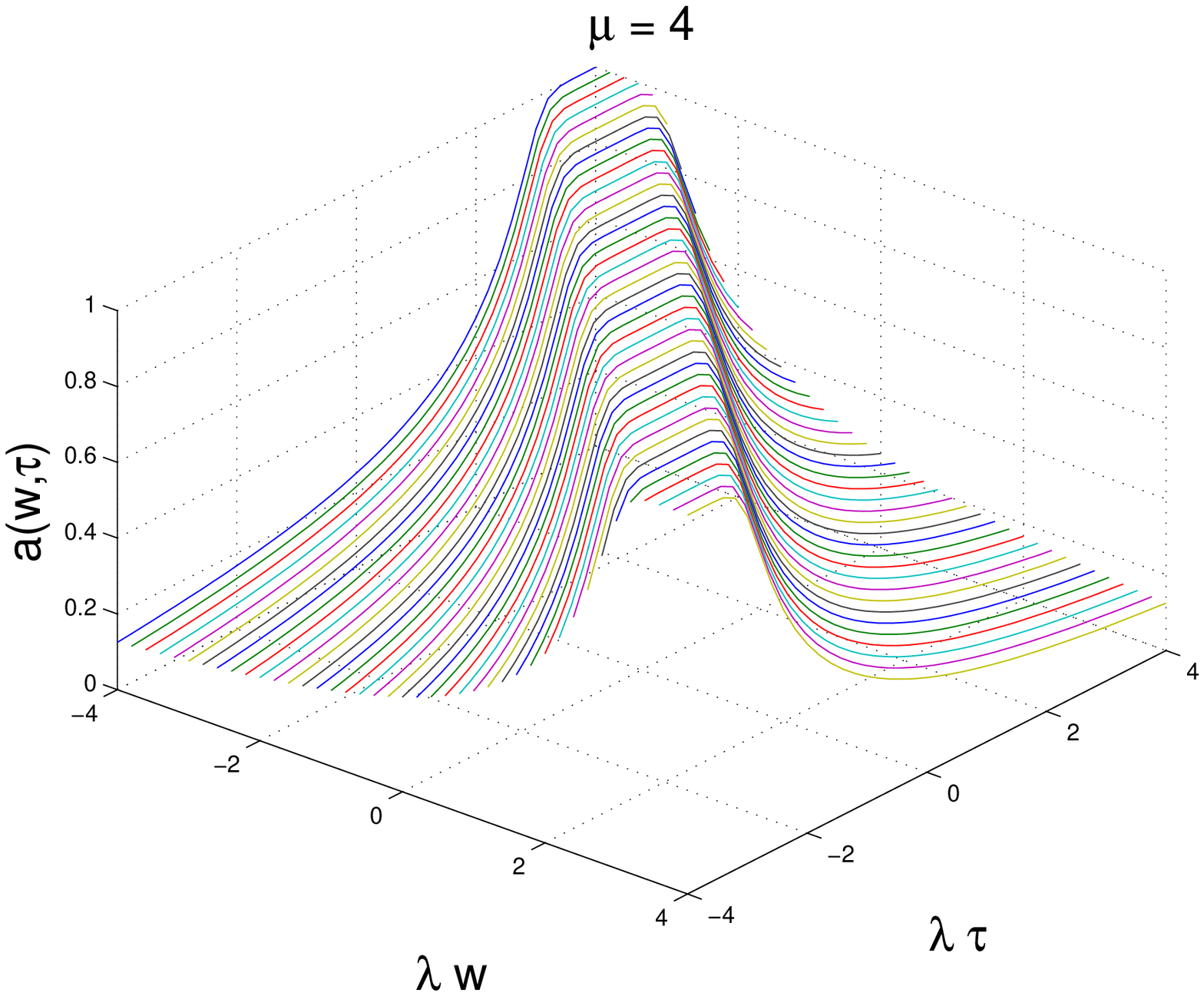}}\\
      \hline
\end{tabular}
\end{center}
\caption[a]{The warp factors for different values of the integer $\mu$.}
\label{F1}
\end{figure}
The solution of Eqs. (\ref{S1}), (\ref{S2}) and (\ref{S3}) can be generalized to a family of solutions 
characterized by an integer parameter $\mu$. The solution can be written as 
\begin{equation}
 a(w,\tau)= \frac{1}{[\lambda^{2\mu} ( w + \tau)^{2\mu} +1]^{\frac{1}{2\mu}}}, \qquad b(w,\tau) = \epsilon \,a(w,\tau),
\label{S1a}
\end{equation}
where, as before, $\epsilon$ is a constant. The form of the warp factor given in  Eq. (\ref{S1a}) solves the whole system provided 
\begin{eqnarray}
&&\varphi(w,\tau) = \sqrt{\frac{3( 2 \mu -1)}{\kappa}} \arctan{[\lambda^{\mu} (w + \tau)^{\mu}]}, 
\label{S2a}\\
&& V(\varphi) =  \frac{3 \lambda^2}{ 2 \kappa} \biggl(\frac{\epsilon^2 -1}{\epsilon^2}\biggr)\Lambda^{\frac{\mu-1}{\mu}}(\varphi) [ (2 \mu + 3) \Lambda(\varphi) - (2\mu -1)],
\label{S3a}
\end{eqnarray}
where $\mu \geq 1$ and  
\begin{equation}
\Lambda(\varphi) = \sin^2\Phi, \qquad \Phi = \sqrt{\frac{\kappa}{3(2 \mu -1)}}\varphi.
\label{S4a}
\end{equation}
In the case $\mu = 1$ the solution expressed by Eqs. (\ref{S1a}), (\ref{S2a}) and (\ref{S3a}) 
goes exactly to the solution of Eqs. (\ref{S1}), (\ref{S2}) and (\ref{S3}). Again, in the limit 
$\epsilon = 1$ Eqs. (\ref{S1a}) and (\ref{S2a}) still solve the system with vanishing potential. 

\begin{figure}
\begin{center}
\begin{tabular}{|c|c|}
      \hline
      \hbox{\epsfxsize = 7.5 cm  \epsffile{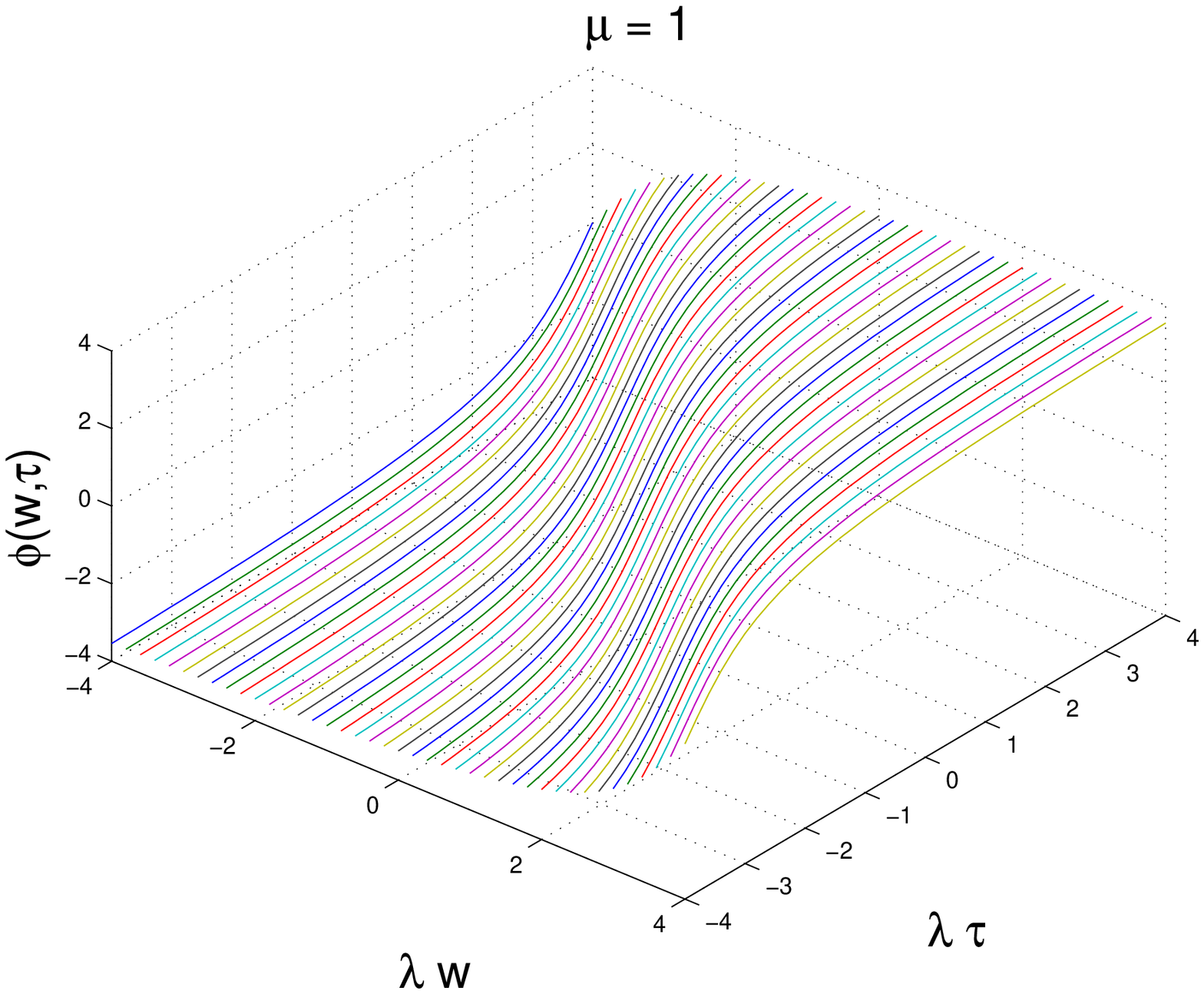}} &
      \hbox{\epsfxsize = 7.5 cm  \epsffile{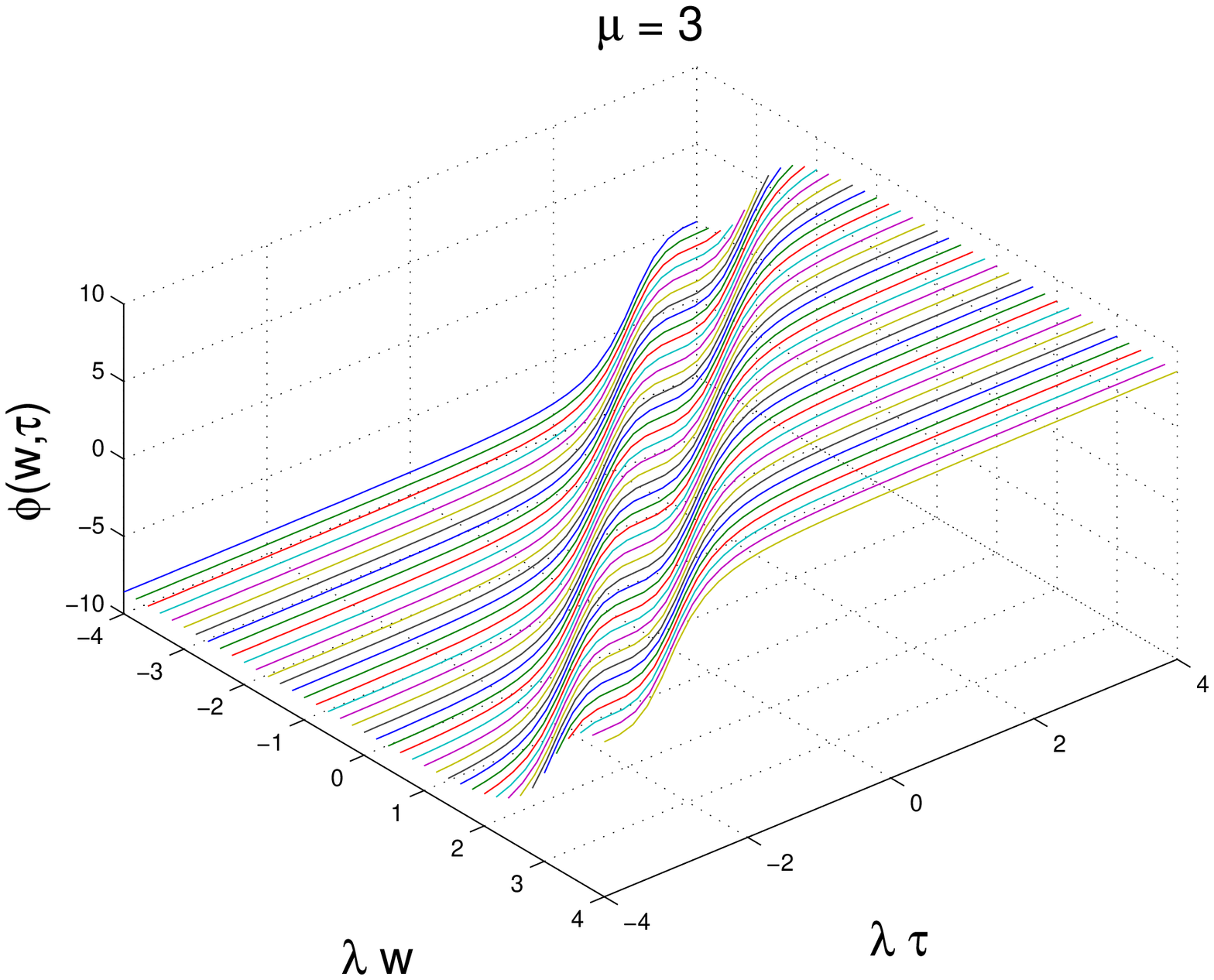}}\\
      \hline
\end{tabular}
\end{center}
\caption[a]{The field profile for odd values of $\mu$.  Natural gravitational units $2 \kappa =1$ 
have been adopted.}
\label{F2}
\end{figure}

\begin{figure}
\begin{center}
\begin{tabular}{|c|c|}
      \hline
      \hbox{\epsfxsize = 7.5 cm  \epsffile{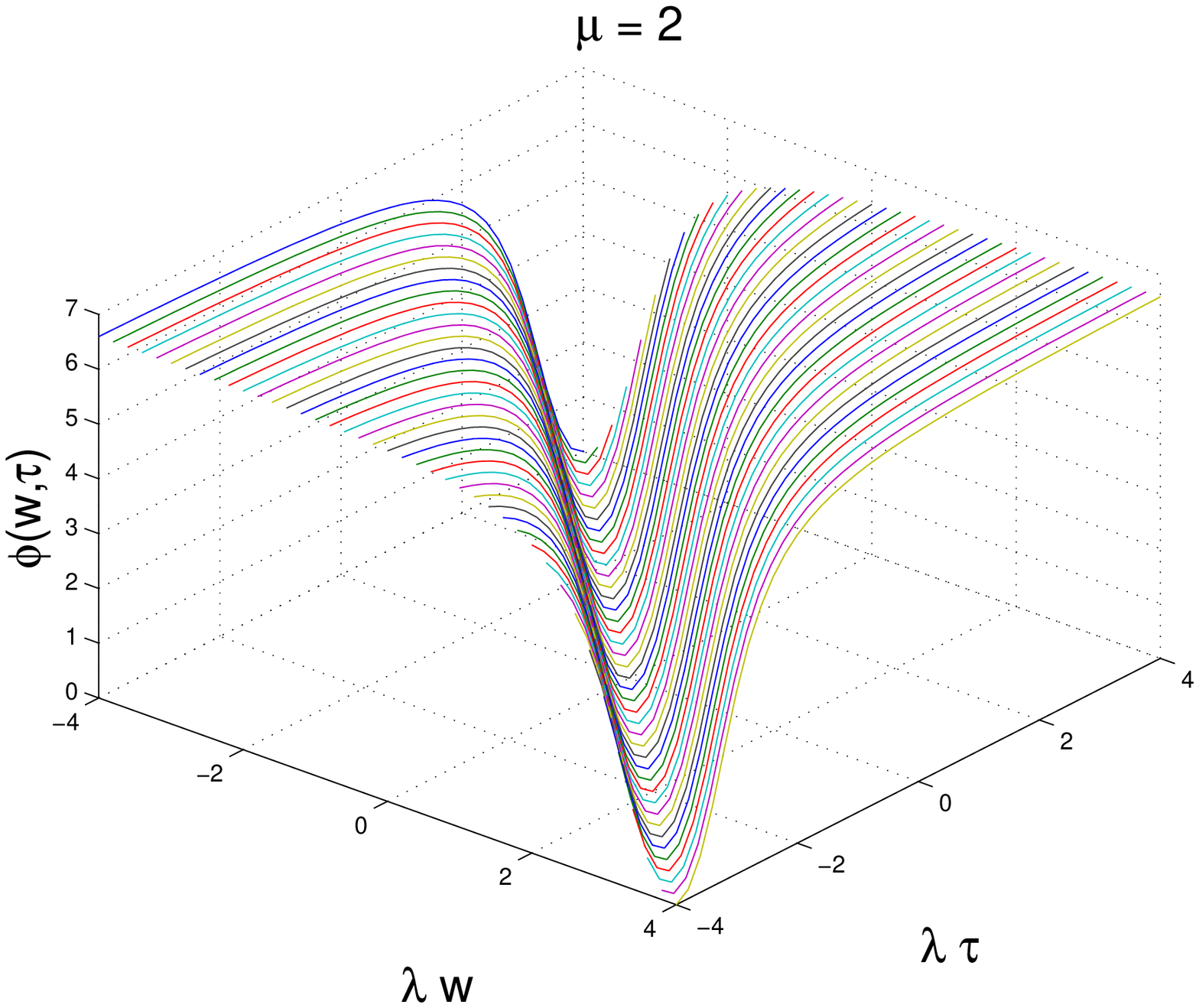}} &
      \hbox{\epsfxsize = 7.5 cm  \epsffile{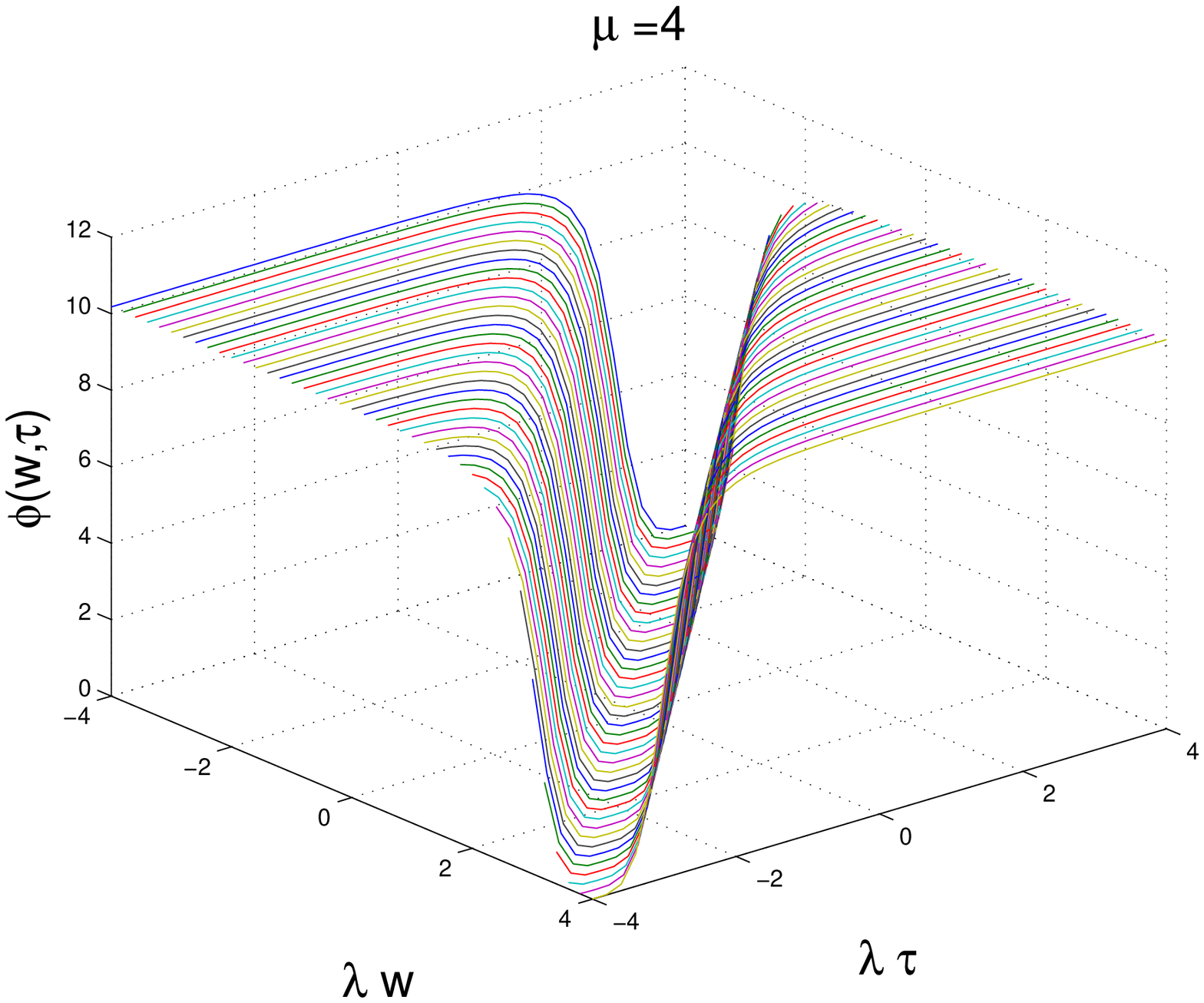}}\\
      \hline
\end{tabular}
\end{center}
\caption[a]{The field profiles for even values of $\mu$ are reported in the natural gravitational units 
introduced in Fig. \ref{F2}.}
\label{F3}
\end{figure}

Different values of $\mu$ imply different properties for the topology of the field solution. 
This point is illustrated in Figs. \ref{F1}, \ref{F2} and \ref{F3}
 where the warp factors and the field profiles are reported for different values of the integer 
 $\mu$ that appears in Eqs. (\ref{S1a}), (\ref{S2a}) and (\ref{S3a}).
According to Fig. \ref{F1}, when $\mu$ increases from $1$ to $4$ 
the warp factor becomes broader in the $(w,t)$ plane (recall that $b(w,\tau) = \epsilon a(w,\tau)$). For fixed value of the conformal 
time coordinate, say $\tau=0$, $a(w,0) \to |\lambda w|^{-1}$ for $|\lambda w|\to \infty$ and $b(w,0)\to 
\epsilon |\lambda w|^{-1}$. In this case the space-time geometry can be viewed, for large absolute 
value of the bulk radius, as a slight deformation of $\mathrm{AdS}_{5}$. The 
deformation parameter is exactly $\epsilon$, which is dimensionless. As soon as $\epsilon \to 1$ the exact 
 $\mathrm{AdS}_{5}$ limit is recovered for $|w|\to \infty$. The geometry is always regular and all 
 the curvature invariants do not have any pole for any finite values of either $w$ or $\tau$. In the limit 
 $\epsilon \to 1$ the Weyl invariant vanishes.
A change in the value of $\mu$ does affect crucially the topological nature 
of the field profile. This aspect can be simply understood from Eq. (\ref{S2a}). If 
$\mu$ is even the field profile interpolates between two different minima of the underlying 
potential and it has the characteristic kink-like shape. In Fig. \ref{F2} this aspect
is illustrated for two different odd values of $\mu$, i.e. $\mu=1$ and $\mu=3$.
Notice a slight difference between the case $\mu=1$ and the case $\mu = 3$.
In the case $\mu=3$ (see Fig. \ref{F2}, plot at the right) the profile flattens around the origin 
of the coordinate system. In fact, in the case $\mu =1$ the first 
derivative of $\Phi(w,\tau)$ (both with respect to $w$ and $\tau$) is not vanishing at the origin, i.e. 
for $w=\tau =0$. On the contrary, at the same point, in the 
case $\mu=3$ the first and second derivatives of $\Phi(w,\tau)$ are both equal to $0$ and the 
only non-vanishing derivative at the origin will be the third one.
Besides this point we can safely say that, when $\mu$ is odd, the field profile 
exhibits a topological nature. Conversely, when $\mu$ is even, the field profile 
does not have a kink-like structure but rather a bag-like structure 
signaling that the corresponding field profile is non-topological.
This aspect is usefully illustrated in Fig. \ref{F3} where $\varphi(w,\tau)$ is reported in the 
cases $\mu = 2$ and $\mu =4$. As it appears from Eq. (\ref{S2a}), when $\mu$ is even, 
the field profile tends to the same constant values when $w$ and $\tau$ are large 
in absolute values.
\begin{figure}
\begin{center}
\begin{tabular}{|c|c|}
      \hline
      \hbox{\epsfxsize = 7.5 cm  \epsffile{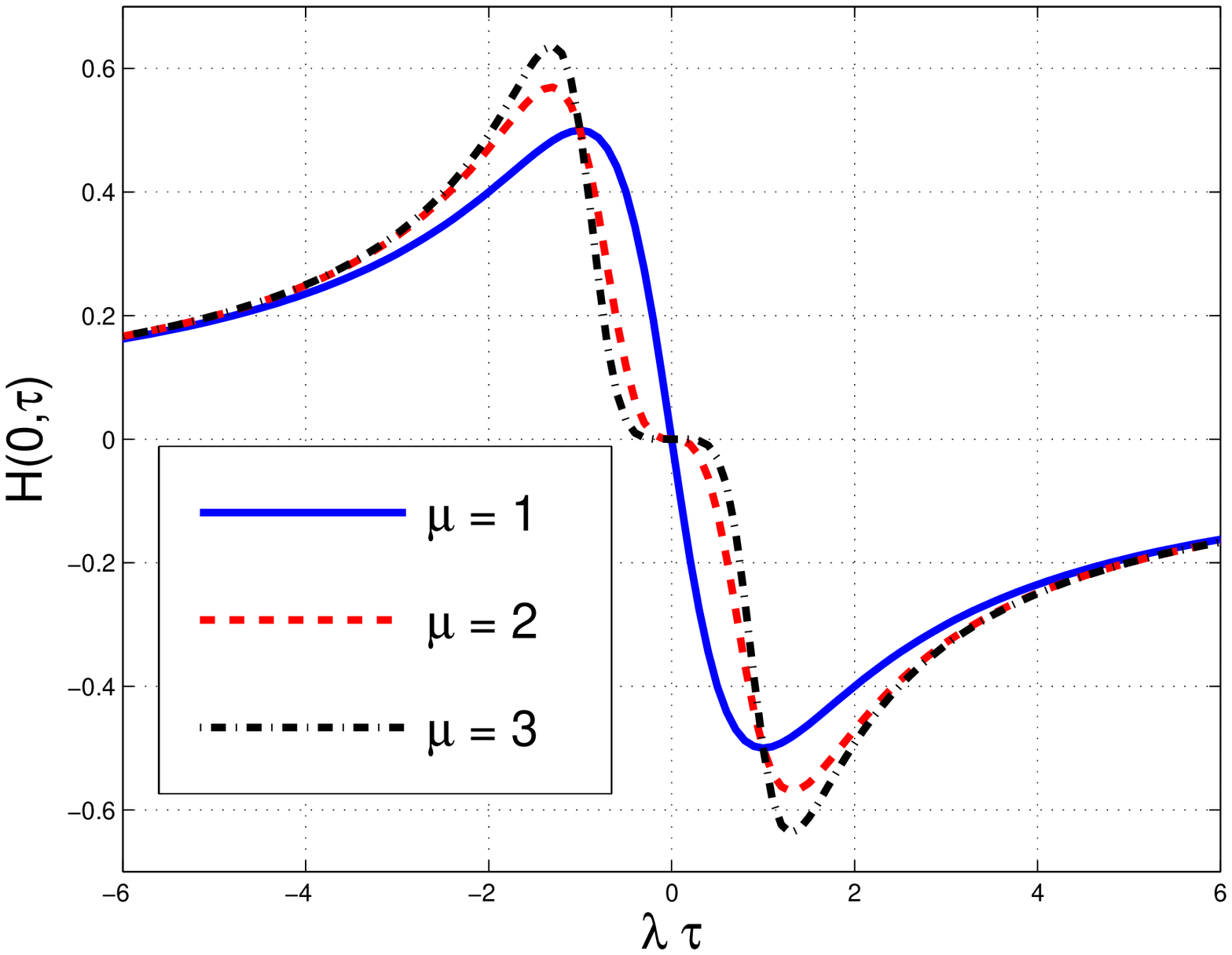}} &
      \hbox{\epsfxsize = 7.5 cm  \epsffile{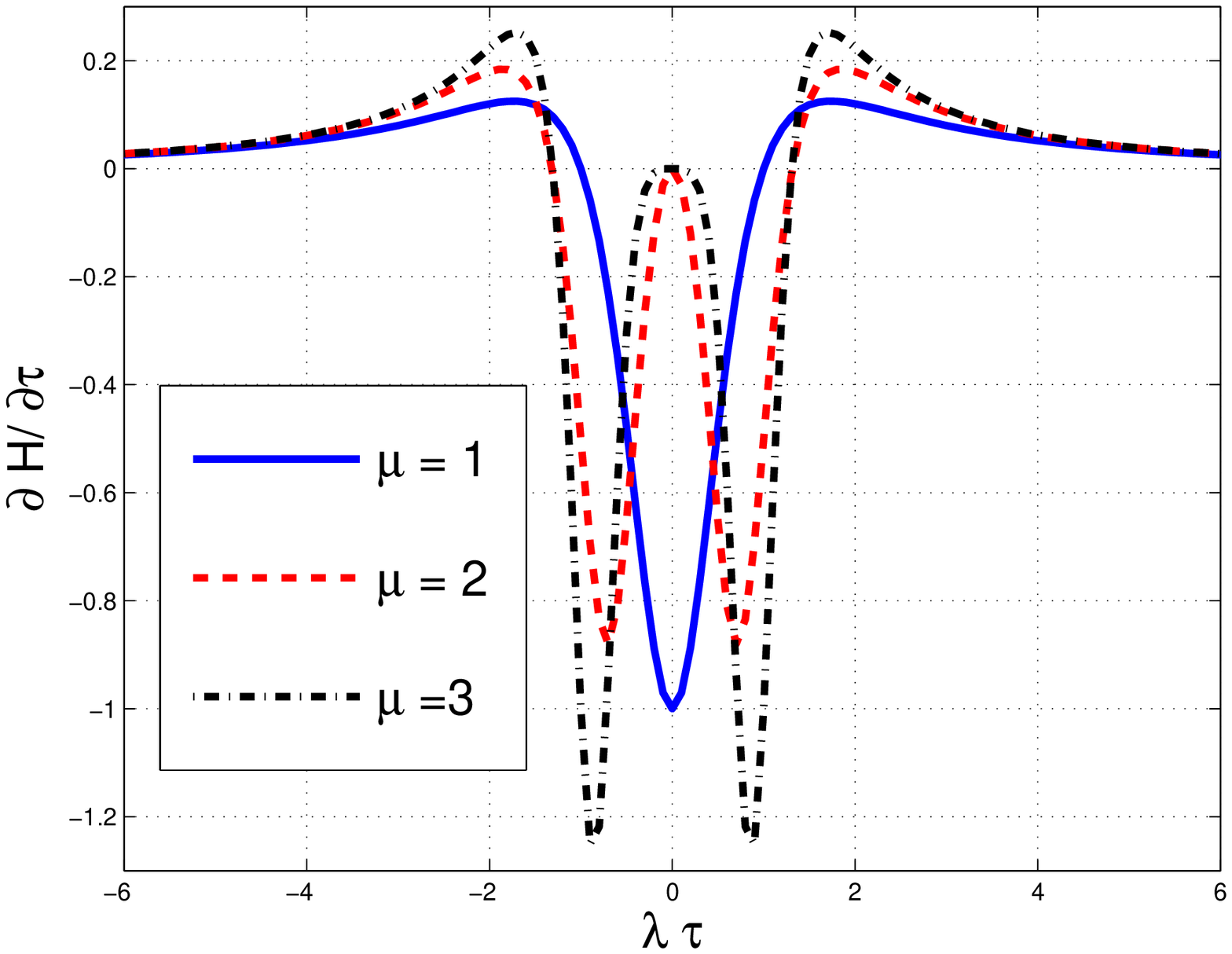}}\\
      \hline
\end{tabular}
\end{center}
\caption[a]{The conformal time evolution of $H$ and $\dot{H}$ is illustrated at a fixed 
value of the bulk radius and for different values of $\mu$.}
\label{F4}
\end{figure}
As far as the cosmological properties of the solutions are concerned, it is useful 
to look at Fig. \ref{F4} where $H$ and $\dot{H}$ are reported as a function of the 
conformal time coordinate $\tau$. It is clear from Fig. \ref{F4} 
that, for fixed value of the bulk radius, say $w =0$, $H(\tau)$ changes 
sign once and $\dot{H}$ changes sign twice. This behaviour is typical of a well known class 
of cosmological scenarios called bounces. Bouncing solutions in the Einstein frame (which is the one 
adopted in the present analysis) are solutions where the Hubble rate flips its sign \cite{39,40} (see also \cite{41}
and references therein).
 The bouncing behaviour obtained 
in the present solutions and illustrated in Fig. \ref{F4} is, at the same time, unrealistic and encouraging.
It is unrealistic since a physical bounce should have features that are opposite to the ones 
exhibited  by Fig. \ref{F4}. In other words a realistic bounce should have $H<0$ for $\tau < 0$ and $ H>0$ for $\tau>0$. This type of evolution would imply that the contraction took place in the far past (for $\tau <0$) while today, as 
we know, the Universe expands rather than contracting. In Fig. \ref{F4} the behaviour of $H$ (and $\dot{H}$)
is opposite: the Universe expands in the past and contracts in the future, which is unrealistic.
At the same time, however, the obtained results are encouraging since we showed, analytically, that 
a bouncing behaviour arises naturally.
 
It is relevant to mention that the only separable solutions that are allowed by (single)
gravitating solitons are just the trivial ones where $H=0$ and $F=0$. To demonstrate this point
consider  the following ansatz for the warp factors:
\begin{equation}
a(w,\tau) = \frac{\alpha(\tau)}{\sqrt{\lambda^2 w^2 + 1}},\qquad b(w,\tau) = \frac{\beta(\tau)}{\sqrt{\lambda^2 w^2 + 1}}.
\label{S1b}
\end{equation}
supplemented by the condition $\varphi = \varphi(w)$.
Using Eq. (\ref{S1b}) into Eqs. (\ref{HFw}) and (\ref{HFtau}) it follows that $\dot{{\mathcal H}}=0$ and $H'=0$. 
But then the momentum constraint of Eq. (\ref{EQ10w}) implies that, since $\dot{\varphi}=0$, 
also $F=0$.  The requirement that the solution is separable demands, moreover, that all 
the pieces containing $H$, $F$ and their derivatives vanish in Eqs. (\ref{EQ100}), (\ref{EQ1ij}), (\ref{EQ1ww}) 
and (\ref{KG1}). This condition translates, after some algebra,
in the following pair of equations
\begin{equation}
\dot{H} +H^2 =0,\qquad HF + H^2=0.
\label{cond1}
\end{equation}
But since $F=0$ (because of the momentum constraint) it must also happen, according to 
Eq. (\ref{cond1}) that $H=0$. Thus, $\alpha(\tau)$ and $\beta(\tau)$ can only be constants. 

\renewcommand{\theequation}{3.\arabic{equation}}
\section{Time-dependent gravitating multi-defects}
\begin{figure}
\begin{center}
\begin{tabular}{|c|c|}
      \hline
      \hbox{\epsfxsize = 7.5 cm  \epsffile{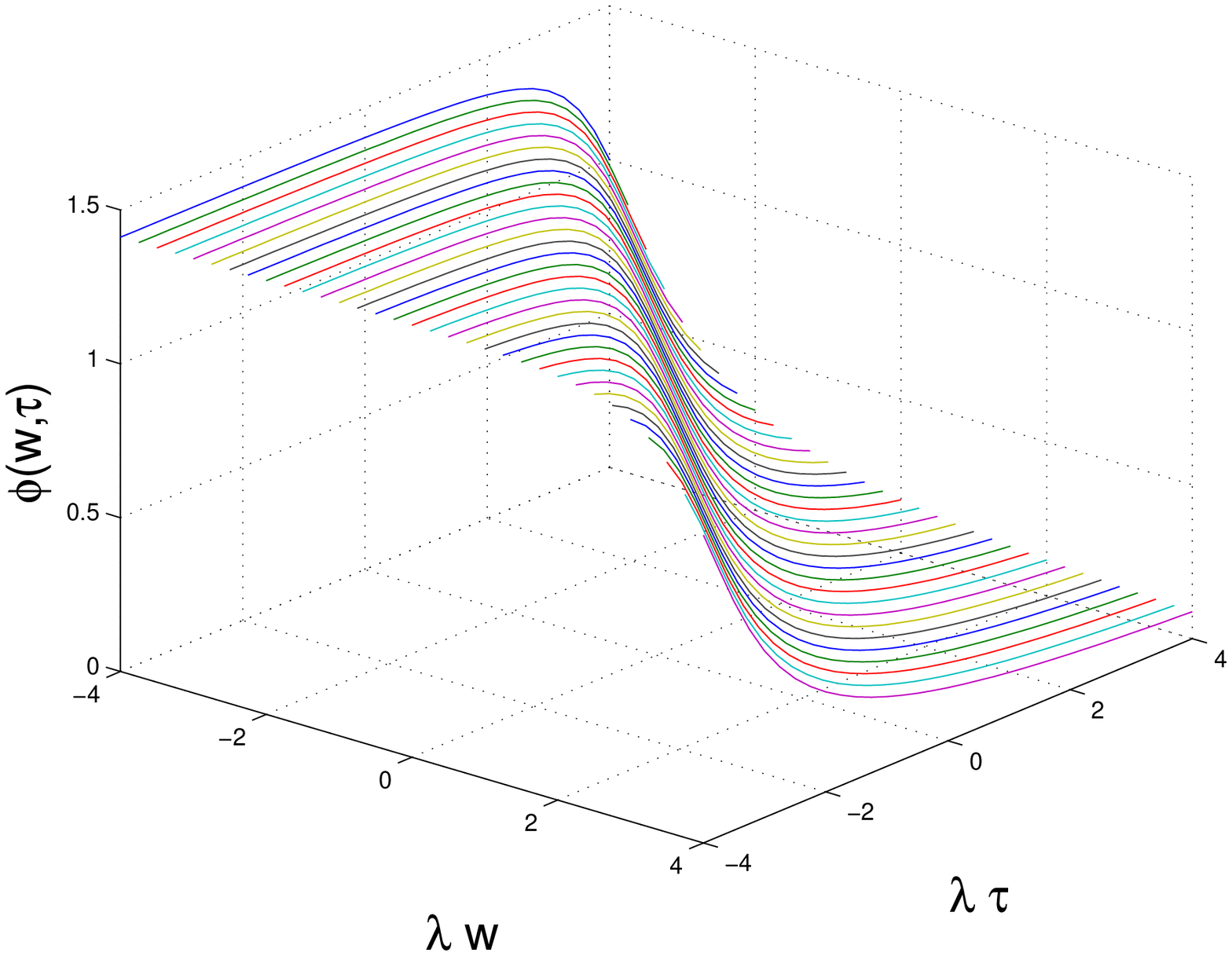}} &
      \hbox{\epsfxsize = 7.5 cm  \epsffile{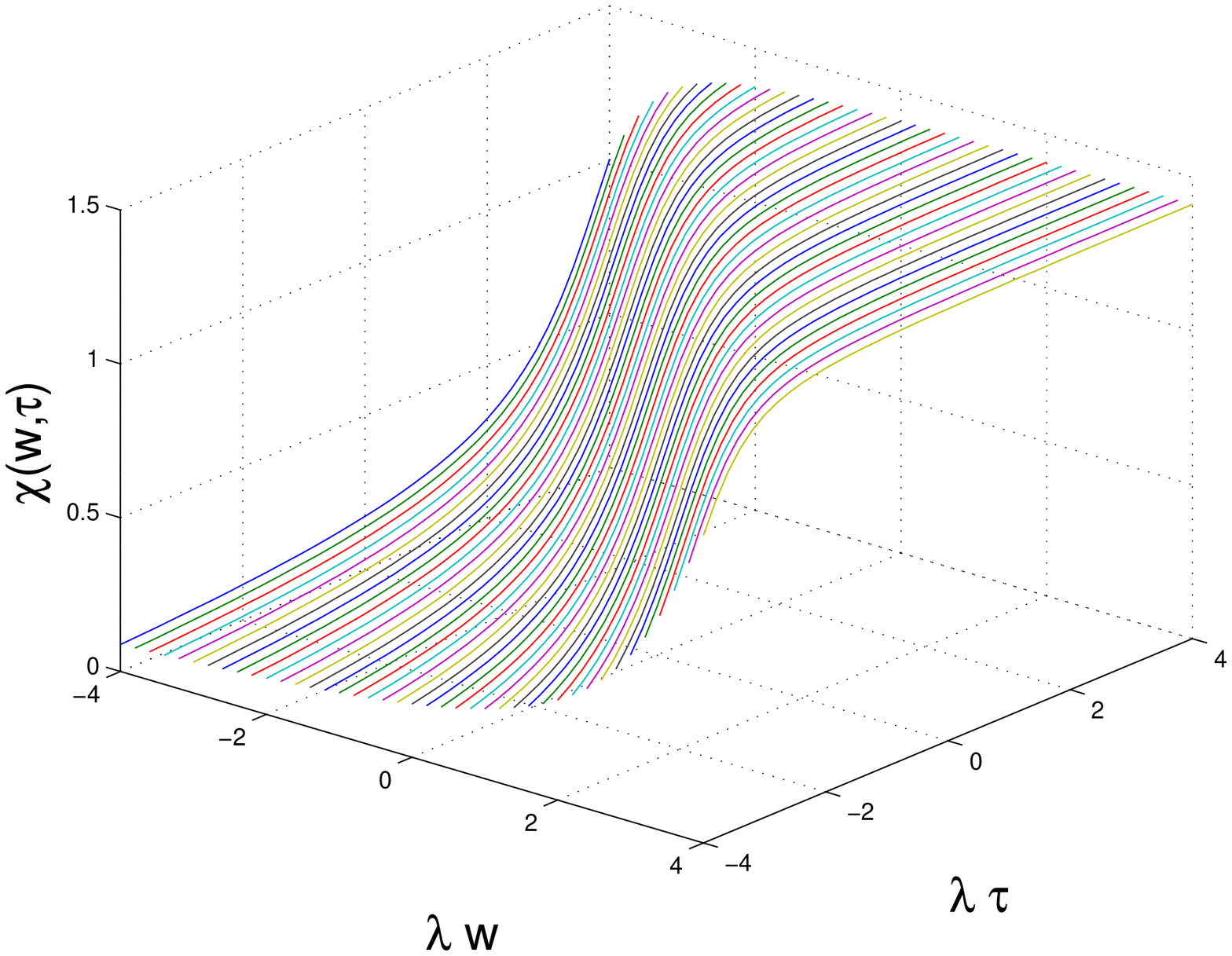}}\\
      \hline
\end{tabular}
\end{center}
\caption[a]{The kink-antikink system of Eq. (\ref{TFan1}) in the case $\tilde{v}=1$ and always 
in natural gravitational units.}
\label{F5}
\end{figure}

\setcounter{equation}{0}
The considerations developed in the previous section can be generalized 
to the situation where two scalar degrees of freedom are simultaneously 
present. The action to be discussed in the present section will therefore be:
\begin{equation}
S = \int d^{5} x \sqrt{|G|} \biggl[ - \frac{R}{2 \kappa} + \frac{1}{2} G^{\mathrm{AB}} \partial_{A}\varphi
\partial_{B}\varphi  + \frac{1}{2} G^{\mathrm{AB}} \partial_{A}\chi
\partial_{B}\chi- W(\varphi,\chi) \biggr].
\label{action2}
\end{equation}
Using the same conventions of the previous section the  evolution equations 
only involving $\varphi$, $\chi$ and their first derivatives 
can be written, in the geometry defined by Eq. (\ref{LEL2}), as 
\begin{eqnarray}
&&\dot{\varphi}^2 + \dot{\chi}^2 = \frac{1}{\kappa} [ 2 H( H + F) - 2 \dot{H} - F^2 - \dot{F}],
\label{TF1}\\
&& \dot{\varphi}\varphi' + \dot{\chi} \chi' = \frac{1}{\kappa} [ \dot{{\mathcal H}} - 4 H' + 3 F {\mathcal H}],
\label{TF2}\\
&& {\varphi'}^2 + {\chi'}^2 = \frac{3}{\kappa} ({\mathcal H} {\mathcal F} - {\mathcal H}') 
+ \frac{b^2}{\kappa a^2}[ (\dot{F} - \dot{H}) + F^2 + HF - 2 H^2].
\label{TF3}
\end{eqnarray}
The remaining three equations involve directly the scalar potential $W(\varphi,\chi)$ and they are:
\begin{eqnarray}
&& W(\varphi,\chi) = \frac{3}{2 \kappa a^2} ( \dot{H} + 2 H^2 + HF) -  \frac{3}{2 \kappa b^2}({\mathcal H}' +
4 {\mathcal H}^2 - {\mathcal H} {\mathcal F}),
\label{TF4}\\
&& \ddot{\varphi} - \frac{a^2}{b^2} \varphi'' + [( 2 H + F) \dot{\varphi} - \frac{a^2}{b^2} (4 {\mathcal H} - {\mathcal F}) 
\varphi'] + a^2 \frac{\partial W}{\partial \varphi} =0,
\label{TF5}\\
&&  \ddot{\chi} - \frac{a^2}{b^2} \chi'' + [( 2 H + F) \dot{\chi} - \frac{a^2}{b^2} (4 {\mathcal H} - {\mathcal F}) 
\chi'] + a^2 \frac{\partial W}{\partial \chi} =0.
\label{TF6}
\end{eqnarray}
Consider, therefore, the following ansatz for the field profiles and for the warp factors:
\begin{eqnarray}
&& \varphi(w,\tau) = \tilde{v} \sqrt{1 + g(w,\tau)}, \qquad \chi(w,\tau) = \tilde{v} \sqrt{1 - g(w,\tau)},
\label{TFan1}\\
&& a(w,\tau) = \sqrt{1 - g^2(w,\tau)}, \qquad b(w,\tau) = \epsilon \, a(w,\tau).
\label{TFan2}
\end{eqnarray}
where $g(w,\tau)$ is a monotonic function which is continuous with its first and second 
derivatives with respect to $w$ and $\tau$. Furthermore, we do assume that 
$|g(w,\tau)|< 1$ so that the square roots that appear ubiquitously in the ansatz are always 
well defined in the domain where $w$ and $\tau$ are allowed to vary. 
By inserting Eqs. (\ref{TFan1}) and (\ref{TFan2})  into Eqs. (\ref{TF1}), (\ref{TF2}) and (\ref{TF3}) 
the following conditions on $g(w,\tau)$ can be explicitly obtained:
\begin{equation}
\dot{g}'  = - \frac{3 g \dot{g} g'}{1 - g^2},\qquad g'' = \dot{g}' \frac{g'}{\dot{g}},\qquad 
\ddot{g} = g'' \frac{\dot{g}^2}{{g'}^2}.
\label{TFan3}
\end{equation}
To obtain Eq. (\ref{TFan3}) the arbitrary constant $v$ has been expressed in terms of $\kappa$ as 
$\kappa V^2 =6$. The solution of Eq. (\ref{TFan3}) becomes then:
\begin{equation}
g(w,\tau) = \frac{\lambda (w + \tau)}{\sqrt{\lambda^2 (w +\tau)^2 + 1}}.
\label{TFan4}
\end{equation}
Equations (\ref{TF4}), (\ref{TF5}) and (\ref{TF6}) can then be used to determine 
the scalar potential that becomes, within our conventions, 
\begin{eqnarray}
W(\varphi,\chi) &=& W_{0}(\epsilon,\tilde{v}) \biggl[ \frac{6}{5} \tilde{v}^2 (\varphi^2 + \chi^2) - (\varphi^4 + \chi^4)\biggr]
\nonumber\\
&+& W_{1} (\epsilon, \tilde{v}) ( \varphi^2 + \chi^2 -2V^2) \biggl[ \frac{11}{8} (\varphi^4 + \chi^4) - \tilde{v}^2 (\varphi^2 + \chi^2)\biggr],
\label{TFan5}
\end{eqnarray}
where 
\begin{equation}
W_{0}(\epsilon,\tilde{v}) = \frac{15 \lambda^2}{4 \kappa \tilde{v}^4} \biggl(\frac{1 - \epsilon^2}{\epsilon^2}\biggr), 
\qquad W_{1}(\epsilon,\tilde{v}) = \frac{\lambda^2}{2 \tilde{v}^4} \biggl(\frac{1 - \epsilon^2}{\epsilon^2}\biggr).
\label{TFan6}
\end{equation}
It can be directly checked that Eqs. (\ref{TFan5}) and (\ref{TFan6}) solve directly Eqs. (\ref{TF4}), (\ref{TF5}) 
and (\ref{TF6}). 
The physical situation described by the reported solution is the one where we have a kink and an anti-kink 
both time dependent. This aspect is illustrated by Fig. \ref{F5} where the functions 
$\varphi(w,\tau)$ and $\chi(w,\tau)$ are separately reported in the plot at the left and in the plot at the right.

As it was shown in the previous section, the one-field system only allows separable solutions 
which are trivial. If two fields are present the situation is different.  Consider, for instance,  
the situation where the warp factors of the line element of Eq. (\ref{LEL2}) are parametrized as 
\begin{equation}
a(w,\tau) = \frac{\alpha(\tau)}{\sqrt{\lambda^2 w^2 +1}},\qquad b(w) = \frac{1}{\sqrt{\lambda^2 w^2 + 1}},
\label{sep1}
\end{equation}
namely $b(w)$ is just static while the time dependence of $a(w,\tau)$ is separable.
In this situation Eq. (\ref{TF2}) is trivially satisfied while Eqs. (\ref{TF1}) and (\ref{TF3}) lead, respectively, 
to the following pair of conditions:
\begin{eqnarray}
&&{\varphi'}^2 = \frac{3}{\kappa} ({\mathcal H}^2 - {\mathcal H}') - \frac{1}{\kappa \alpha^2}[\dot{H} + 2 H^2],
\label{sep2}\\
&& \dot{\chi}^2 = \frac{2}{\kappa} (H^2 - \dot{H}).
\label{sep3}
\end{eqnarray}
The potential can then be written as 
\begin{equation}
W(\varphi,\chi) = \frac{3}{2 \kappa a^2}(\dot{H} + 2 H^2)- \frac{3}{2 \kappa b^2}({\mathcal H}' + 3 {\mathcal H}^2).
\label{sep4}
\end{equation}
There are now different possibilities and, for purposes of illustration, let us focus the attention on the 
simplest one, namely the case where $\dot{H} + 2 H^2 =0$. In this situation the potential will be effectively 
independent on $\chi$ and the evolution of $\alpha(\tau)$ will be simply 
driven by the kinetic energy of $\chi$. In fact, Eqs. (\ref{sep2}) and (\ref{sep3}) reduce to 
\begin{eqnarray}
&& \varphi' = \sqrt{\frac{3}{\kappa}} \frac{\lambda}{\lambda^2 w^2 +1},
\label{sep5}\\
&& \dot{\chi} = \sqrt{\frac{6}{\kappa}} H(\tau)
\label{sep6}
\end{eqnarray}
and $H(\tau)= (2 \tau)^{-1}$ which is, as expected, the expansion rate dictated typical of a 
FRW Universe filled by  a free (minimally coupled) scalar degree of freedom. 
Since $\dot{H} = - 2 H^2$, Eq. (\ref{sep4}) allows to determine also the potential 
of $\varphi$. The scalar profile $\varphi(w)$ and its related potential coincide, therefore, with 
the ones reported, respectively, in Eqs. (\ref{SOL1}) and (\ref{SOL2}).

\renewcommand{\theequation}{4.\arabic{equation}}
\section{Traveling solitons}
\setcounter{equation}{0}

In this section we are going to discuss the case of traveling soliton solutions. More specifically 
it will be shown that, under certain conditions, every static solution cane be promoted, by Lorentz 
invariance, to traveling soliton solution. To discuss this point it is useful to write 
some of the equations of the previous section in terms of the new boosted coordinates:
\begin{equation}
u = \frac{w + U \tau}{\sqrt{1 - U^2}},\qquad v = \frac{w - U \tau}{\sqrt{1 - U^2}}.
\label{Lcone}
\end{equation}
Consider therefore the line element of Eq. (\ref{LEL2}) with $a(w,\tau)= b(w,\tau)$.
After few linear combinations it is easy to show, with simple algebra, that Eqs. (\ref{TF1}), (\ref{TF2}) 
and (\ref{TF3}) translate in the following triplet of equations:
\begin{eqnarray}
&&\varphi_{u}^2 + \chi_{u}^2 = \frac{3}{\kappa}(h_{u}^2 - h_{uu}),
\label{LC1}\\
&& \varphi_{v}^2 + \chi_{v}^2 = \frac{3}{\kappa}(h_{v}^2 - h_{vv}),
\label{LC2}\\
&& \varphi_{u} \varphi_{v} + \chi_{u} \chi_{v} = \frac{3}{\kappa}(h_{u} h_{v} - h_{uv}),
\label{LC3}
\end{eqnarray}
where the subscript denotes a partial derivation with respect to the corresponding variables; moreover 
$h = \ln{a}$.
Equation (\ref{TF4}) implies, in terms of the variables o Eq. (\ref{Lcone}), 
\begin{equation}
W(\varphi,\chi) = - \frac{3}{2\kappa} [( h_{uu} + h_{vv} + 3 h_{u}^2 + 3 h_{v}^2) 
+ 2 \cosh{2\alpha} (h_{uv} + 3 h_{u} h_{v})],
\label{LC4}
\end{equation}
where $\cosh{2\alpha} = (1 + U^2)/(1 - U^2)$.
It is finally easy to show that Eqs. (\ref{TF5}) and (\ref{TF6}) lead, respectively, to the following 
pair of equations:
\begin{eqnarray}
&& \varphi_{uv} - \varphi_{uu} - \varphi_{vv} - 3 (h_{u} \varphi_{u} + h_{v} \varphi_{v}) - 3 \cosh{2\alpha} (h_{u} \varphi_{v} + h_{v} \varphi_{u}) + a^2 \frac{\partial W}{\partial\varphi}=0,
\label{LC5}\\
&& \chi_{uv} - \chi_{uu} - \chi_{vv} - 3 (h_{u} \chi_{u} + h_{v} \chi_{v}) - 3 \cosh{2\alpha} (h_{u} \chi_{v} 
+ h_{v} \chi_{u}) + a^2 \frac{\partial W}{\partial\chi}=0,
\label{LC6}
\end{eqnarray}
Suppose now that both, $\varphi$ and $\chi$ are functions of $u$, i.e. 
$\varphi= \varphi(u)$ and $\chi = \chi_{u}$. Then, Eqs. (\ref{LC2}) and (\ref{LC3}) 
will be satisfied provided $h = h(u)$, i.e. $a= a(u)$. Under these conditions 
the relevant set of equations to be solved becomes:
\begin{eqnarray}
&&\varphi_{u}^2 + \chi_{u}^2 = \frac{3}{\kappa}(h_{u}^2 - h_{uu}),
\label{EX1}\\
&& W(\varphi,\chi) = - \frac{3}{2\kappa} (h_{uu} + 3 h_{u}^2)
\label{EX2}\\
&& \varphi_{uu} + 3 h_{u} \varphi_{u}  -  a^2 \frac{\partial W}{\partial\varphi}=0,
\label{EX3}\\
&& \chi_{uu} +  3 h_{u} \chi_{u}  -  a^2 \frac{\partial W}{\partial\chi}=0.
\label{EX4}
\end{eqnarray}
It can be easily seen that Eqs. (\ref{EX1}), (\ref{EX2}), (\ref{EX3}) and (\ref{EX4}) 
have the same form of the static solutions (in the two field case) when $a(w,\tau)= b(w,\tau)$.
The derivative with respect to $w$ translates into a derivative with respect to $u$.
So, every static solution can be boosted and the resulting metric and field profiles will still be a 
solution. The same conclusion can be reached if the boost is performed in the 
opposite direction. In this second case the field variables will only be functions of $v$ (and not 
of $u$) and the relevant equations can be obtained from Eqs. (\ref{EX1})--(\ref{EX4}) 
by replacing $u \to v$ in the derivatives of $\phi$, $\chi$ and $h$. 

As an example it is useful to consider the following solution of the above field equations in the 
case where, both field profiles are only functions of, say, $u$:
\begin{eqnarray}
&& \varphi(u) = \varphi_{0} \{ [1 + q(u)]^{3/2} + [ 1 - q(u)]^{3/2}\},
\nonumber\\
&& \chi(u) = \chi_{0}  \{ [1 + q(u)]^{3/2} - [ 1 - q(u)]^{3/2}\},
\label{EX5}
\end{eqnarray}
where $q(u)$ is a continuous and differentiable function of $u$; $\varphi_{0}$ and $\chi_{0}$ are both constants.
 As the function $g(w,\tau)$ 
introduced in the previous section, it will be required that $|q(u)|<1$ in the whole 
domain of definition. To satisfy the constraint of Eq. (\ref{LC3}) it will be necessary that 
also $ a= a(u)$ and the static solution implies that the appropriate ansatz is 
nothing but 
\begin{equation}
a(u) = \frac{1}{(\lambda^{2\mu} u^{2\mu} + 1)^{\frac{1}{2\mu}}}.
\label{EX6}
\end{equation}
The function $q(u)$ is therefore determined to be 
\begin{equation}
q(u) = \frac{2}{\pi} \arctan{[\lambda^{\mu} u^{\mu}]},\qquad \kappa \varphi_{0}^2 = \frac{\pi^2}{12} \frac{2 \mu - 1}{\mu^2}
\label{EX7}
\end{equation}
with $\chi_{0} = \varphi_{0}$. Equations (\ref{EX2}), (\ref{EX3}) and (\ref{EX4}) allow to determine the 
potential:
\begin{equation}
W(\varphi,\chi) = \frac{3 \lambda^2}{2 \kappa} [\sin^2{\sigma}]^{\frac{\mu -1}{\mu}} 
[ ( 2 \mu -1) - (2 \mu + 3) \sin^2{\sigma}] + [|\tilde{\varphi} + \tilde{\chi}|^{2/3} + |\tilde{\varphi} - \tilde{\chi}|^{2/3} -2] {\mathcal A},
\label{POTnu}
\end{equation}
where, as usual, $\\varphi_{0}arphi = 2 \varphi_{0} \tilde{\varphi}$ and $\chi = 2 \varphi_{0} \tilde{\chi}$. The functions 
$\sigma(\tilde{\varphi},\tilde{\chi})$ and ${\mathcal A}(\tilde{\varphi},\tilde{\chi})$ appearing in Eq. (\ref{POTnu}) 
are defined, respectively, as:
\begin{eqnarray}
&&{\mathcal A}(\tilde{\varphi}, \tilde{\chi}) = \frac{18 b^2\, \varphi_{0}^2 \,\mu}{\pi^2} \cos{\sigma}
 \{ \sigma [\sin^2{\sigma}]^{\frac{\mu-2}{2\mu}} [ (\mu -1) - ( 2 \mu + 3) \sin^2{\sigma}] 
\nonumber\\ 
&&+ \frac{\mu}{2} \cos{\sigma} [\sin^2{\sigma}]^{\frac{\mu -1}{\mu}}\},
\label{Anu}\\
&&\sigma(\tilde{\varphi},\tilde{\chi}) = \frac{\pi}{4} [ |\tilde{\varphi} + \tilde{\chi}|^{2/3} - |\tilde{\varphi} - \tilde{\chi}|^{2/3}].
\label{sigmanu}
\end{eqnarray}
As expected this traveling solution has a static analog that has been discussed in \cite{38}.

To conclude the present section let us consider the possibility of having solutions 
that mix $u$ and $w$. There are, of course, trivial solutions to this problem, namely the 
ones where the dependence of the various fields and of the warp factors 
depends either upon $(u + v)$ or upon $(u-v)$. In these two cases the resulting 
profiles are either completely static or completely time-dependent. 

The kind of mixed solutions which would be interesting to obtain would be the five-diemensional 
analog of the $(1+1)$ dimensional solutions of the Sine-Gordon system that have been introduced 
in \cite{1,2} and extensively discussed in \cite{3}.
Indeed, it is well known that there are no static two-kink solutions  in the Sine-Gordon system when
only one scalar degree of freedom is present. However, two-kink solutions in which either kinks have 
arbitrary velocity (but at least one non-zero) can be constructed \cite{1,2,3}. A typical example 
of this dynamics is the field profile given, in $(1+1)$ dimensions, by $4 \arctan{[U \sinh{( \gamma x)}/
\cosh{(\gamma U t)}]}$ where $\gamma= (1 - U^2)^{-1/2}$ and the two relevant 
coordinates have been denoted by $x$ (space) and $t$ (time). In this solution the kink approach the $x$ axis 
with velocity $U$, scatter elastically at $t =0$ and emerge from the interaction with a computable 
phase shift. Notice also that, for $t\to +\infty$ and $x\to +\infty$ the field profile becomes solely 
function of $u= \gamma ( x + U t)$. Conversely, for $x \to +\infty$ and $t \to -\infty$, the 
field profile becomes, asymptotically, only function of $v = \gamma( x - U t)$. 

In the five-dimensional warped geometries discussed in the present paper these solutions are rather hard 
to find and we cannot report any positive result. The main problem is represented by the structure 
of Eqs. (\ref{LC1}), (\ref{LC2}) and (\ref{LC3}).  Indeed it seems rather difficult to satisfy 
simultaneously the momentum constraint of Eq. (\ref{LC3}) together with the conditions 
implied by Eqs. (\ref{LC1}) and (\ref{LC2}). In spite of this negative 
result it seems to be worthwhile to investigate further the possible solutions describing the 
elastic scattering of solitons in five dimensions. These solutions might have an intriguing 
cosmological interpretation.

\renewcommand{\theequation}{5.\arabic{equation}}
\section{Concluding remarks}
\setcounter{equation}{0}

In the present paper time-dependent solitonic solutions have been discussed. It has been shown 
that these solutions exist both for the case of topological solitons and for the case of non-topological 
solitons. The discussion has been also generalized to include the possibility 
of multi-defects, i.e. the situation where more than one defect is present in the system. 
In many respects the results of the present analysis are encouraging but still preliminary. 
They are encouraging since there exist solitonic profiles leading to time-dependent 
geometries that are diagonal and that have a well defined $\mathrm{AdS}_{5}$ limit 
for large values of the bulk radius and for a fixed value of the conformal time coordinate.
The reported results suggests several lines of possible developments.
In the first place more realistic bouncing behaviours may be studied always 
in the case of five-dimensional warped geometries. Moreover, it is not excluded 
that different classes of solutions (like the ones related to elastic scattering 
of two profiles) may emerge.

The main aim of the present script has been to provide examples of thick defects 
that lead to solutions whose time-dependent evolution may be potentially relevant 
for cosmology. In this respect, an obvious extension 
of the reported considerations will be to study higher dimensional defects, like, 
for instance, Abelian vortices in six-dimensions. These models are known 
to provide, at the static level,  $\mathrm{AdS}_{6}$ geometries 
for large values of the bulk radius. 

It would be finally interesting to compute the production of relic gravitons. 
Indeed, while at the static level the zero modes of the gravitons will certainly be localized owing to the 
$\mathrm{AdS}_{5}$ nature of the geometry, if the 
solutions exhibit time-dependence the natural question to ask will be how many gravitons 
are produced by the time-dependent action of the gravitational field. This analysis  
requires necessarily a more realistic construction: it would not be enough, in the present approach,
 to patch different solutions since a more realistic implementation 
 of the bouncing behaviour is required. All these issues are beyond the aims of the present investigation.

\section*{Acknowledgments}
It is a pleasure to thank M. Halpern for valuable discussions.

\end{document}